\documentclass[aps,pra,superscriptaddress,preprint,amsmath,amssymb,showpacs,floatfix]{revtex4-1}
\usepackage{times}
\usepackage[utf8x]{inputenc}
\usepackage[english]{babel}
\usepackage[T1]{fontenc}
\usepackage{lmodern}
\usepackage{bbold}
\usepackage[pdftex]{graphicx}
\usepackage{epstopdf}
\usepackage{bm}
\usepackage{color}
\usepackage{enumerate}

\DeclareMathOperator{\Tr}{Tr}
\begin{document}
\title{Effect of quantum coherence on Landauer's principle}
\author{Kazunari Hashimoto}
\affiliation{
Faculty of Engineering, University of Yamanashi, 4-3-11 Takeda, Kofu, Yamanashi 400-8511, Japan
}
\email{hashimotok@yamanashi.ac.jp}
\author{Chikako Uchiyama}
\affiliation{
Faculty of Engineering, University of Yamanashi, 4-3-11 Takeda, Kofu, Yamanashi 400-8511, Japan
}
\affiliation{
National Institute of Informatics, 2-1-2 Hitotsubashi, Chiyoda-ku, Tokyo 101-8430, Japan
}
\begin{abstract}
Quantum Landauer's principle provides a fundamental lower bound for energy dissipation occurred with information erasure in the quantum regime.
While most studies have related the entropy reduction incorporated with the erasure to the lower bound~(entropic bound), recent efforts have also provided another lower bound associated with the thermal fluctuation of the dissipated energy~(thermodynamic bound).
The coexistence of the two bounds has stimulated comparative studies of their properties; however, these studies were performed for systems where the time-evolution of diagonal (population) and off-diagonal (coherence) elements of the density matrix are decoupled.
In this paper, we aimed to broaden the comparative study to include the influence of quantum coherence induced by the tilted system--reservoir interaction direction.
By examining their dependence on the initial state of the information-bearing system, we find that the following properties of the bounds are generically held regardless of whether the influence of the coherence is present or not:
the entropic bound serves as the tighter bound for a sufficiently mixed initial state, while the thermodynamic bound is tighter when the purity of the initial state is sufficiently high.
The exception is the case where the system dynamics involve only phase relaxation;
in this case, the two bounds coincide when the initial coherence is zero; otherwise, the thermodynamic bound serves the tighter bound. 
We also find the quantum information erasure inevitably accompanies constant energy dissipation caused by the creation of system--reservoir correlation, which may cause an additional source of energetic cost for the erasure.
\end{abstract}

\maketitle
\section{Introduction}
Information processing accompanies inevitable energy dissipation.
According to Landauer's principle~\cite{Landauer61}, the ultimate source of energy dissipation is information erasure, and it is bounded from below by the corresponding reduction in informational entropy.
The principle establishes a fundamental link between information theory and thermodynamics~\cite{Penrose70, Bennet73, Landauer91, Plenio01}.
In the classical regime, its validity has been proven for a wide range of systems theoretically~\cite{shizume95, piechosinska00} and experimentally~\cite{toyabe10, orlov12, berut12, jun14}.
In recent years, rapid developments in quantum technologies stimulate generalizations of the principle to the quantum regime~\cite{piechosinska00, hilt11, reeb14}.
Remarkably, in Ref.~\cite{reeb14}, Reeb and Wolf provided a clear framework for quantum information erasure and successfully derived a quantum version of Landauer's principle, which states that the energy dissipation occurred with the quantum information erasure is lower-bounded by the corresponding reduction of the von Neumann entropy of the information-bearing system.
Since energy dissipation is ubiquitous in quantum operations, its clear understanding is intrinsically important both from fundamental and practical viewpoints.
In this regard, several studies have examined the lower bound for the energy dissipation in quantum information processing~\cite{sagawa09, faist15, mohammady16, bedingham16, peterson16, chitambar19} or operation of quantum heat engine~\cite{goold16review, millen16}.

Despite the conventional Landauer's principle being rooted in the information theory, recent studies in quantum thermodynamics provide another lower bound related to the thermal fluctuation of the dissipated energy~\cite{goold15, guarnieri17}.
Because of its completely different physical origin from the entropic lower bound, subsequent comparative studies on the relative tightness of the two bounds have been stimulated~\cite{goold15, guarnieri17, campbell17, hashimoto20}.
In these studies, the two bounds are compared for systems where a single two-level system transversally contacts with finite~\cite{goold15, guarnieri17, campbell17} or infinite~\cite{hashimoto20} reservoirs.
For the transversal interaction, the dynamics of the population and the coherence are decoupled.
Under the assumption, they clarified the following generic features:
for the systematically changed initial state of the system, the thermodynamic bound depends only on the initial population, whereas the entropic bound is relevant to the initial coherence.
Since the interplay between the population and the coherence is one of the most significant aspects of quantum operations, it is highly desirable to extend the studies to a more generic system--reservoir interaction.
Indeed, a number of recent studies~\cite{miller20, vu22} address the influence of quantum coherence on energy dissipation by evaluating the entropic bound in the presence of longitudinal system--reservoir interaction.
Our main aim in the present paper is to proceed in this direction to the comparative study of the two bounds.

In this paper, we provide a systematic study of the relative tightness of the bounds for the spin--boson model consisting of a single spin-$1/2$ and an infinitely large bosonic reservoir with a tilted system--reservoir interaction direction.
By adjusting the angle of the interaction direction, we control the coupling between the population and the coherence.
Our analysis is based on the full-counting statistics (FCS) formalism of the bounds~\cite{guarnieri17} with the time-convolutionless type quantum master equation, which is time-local even beyond the Markov approximation~\cite{uchiyama14,guarnieri16,hashimoto20,hashimoto19}.
With this formalism, we show that the above-summarized trends of the bounds reported in Ref.~\cite{hashimoto20} hold even under the influence of quantum coherence.
We also point out that Reeb and Wolf's quantum information erasure protocol inevitably accompanies constant energy dissipation caused by the creation of system--reservoir correlation, which may cause an additional source of energetic cost for the erasure.

\section{Thermal quantum information erasure}\label{sec:protocol}
In the original work Ref.~\cite{Landauer61}, R.~Landauer argued to ``erase'' or ``reset'' a classical bit by interacting it with a ``thermal reservoir'' or ``energy sink'', and bringing it to a ``definite'' state.
In the quantum regime, a general framework of the information erasure was formulated in Ref.~\cite{reeb14}, which satisfies the following prerequisites:
\begin{enumerate}
\item the protocol involves an information-bearing system S and a thermal reservoir R, both described by certain Hamiltonians, denoted $H_{S}$ and $H_{R}$, respectively,
\item the reservoir R is initially in the thermal equilibrium with a certain inverse temperature $\beta$, $\rho_{R}(0)=\rho^{{\rm eq}}_{R}\equiv\exp(-\beta H_{R})/\Tr_{R}[\exp(-\beta H_{R})]$, where $\rho_{R}(t)$ is the reduced density operator of R,
\item the system S and the reservoir R are initially uncorrelated, $\rho_{\rm tot}(0)=\rho_{S}(0)\otimes\rho^{{\rm eq}}_{R}$, where $\rho_{\rm tot}(0)$ is the total density operator of S$+$R and $\rho_{S}(t)$ is the reduced density operator of S,
\item the erasure process itself proceeds by a unitary evolution generated by the total Hamiltonian $H=H_{{\rm S}}+H_{R}+H_{SR}$, where $H_{SR}$ is an interaction between S and R.
\end{enumerate}
Following the above framework, we consider a specific protocol of a quantum information erasure: we erase an information content of a spin S by interacting it with an infinite bosonic reservoir R until it reaches a steady-state satisfying $d\rho_{S}(t)/dt=0$.

\section{Lower bounds for the energy dissipation}
The above-formulated information erasure accompanies unavoidable energy exchange, or ``energy dissipation'', between the system and the reservoir.
The actual amount of the dissipated energy can be evaluated as 
	\begin{equation}
  	\label{def:heat}
	\langle\Delta Q\rangle=\Tr_{R}[H_{R}(\rho_{R}(t)-\rho_{R}(0))].
	\end{equation}
Landauer's principle claims that the dissipated energy has a lower bound, meaning that the information erasure requires a specific energetic cost, and it may not be zero.
In the present paper, we systematically compare two lower bounds with different physical origins: (a) the entropic bound defined by the entropy change during the erasure process and (b) the thermodynamic bound defined by the thermal fluctuation of the dissipated energy.
Let us briefly review each bound in the rest of the present section.

\subsection{Entropic bound}
In Ref.~\cite{esposito10, reeb14}, an equality for the dissipated energy $\langle\Delta Q\rangle$ was derived
	\begin{equation}
	\beta\langle\Delta Q\rangle=\Delta S+I(S';R')+D(\rho_{R}(t)||\rho_{R}(0)),
	\end{equation}
where $\Delta S\equiv S(\rho_S(0))-S(\rho_S(t))$, with von Neumann entropy $S(\rho_{S})\equiv-{\rm Tr}_{S}[\rho\ln\rho_{S}]$, is the entropy decrease in the system, $I(S';R')\equiv S(\rho_{S}(t))+S(\rho_{R}(t))-S(\rho_{\rm tot}(t))$ is the mutual information between S and R, quantifying the correlation building up between S and R, and $D(\rho_{R}(t)||\rho_{R}(0))\equiv{\rm Tr}_{E}[\rho_{R}(t)\ln\rho_{R}(t)]-{\rm Tr}_{R}[\rho_{R}(t)\ln\rho_{R}(0)]$ is the relative entropy in R representing the increase in free energy in the environment \cite{esposito10}.
Because any deviation from the initial preparation of the total system, the second and third prerequisites, create a system--reservoir correlation or free energy in the environment, both $I(S';R')$ and $D(\rho_{R}(t)||\rho_{E}(0))$ are positive in the quantum information erasure process \cite{esposito10,reeb14}.
The equality thus provides the quantum version of Landauer's inequality
	\begin{equation}
	\beta\langle\Delta Q\rangle\geq\Delta S,
	\end{equation}
which states that the dissipated energy Eq.~\eqref{def:heat} is bounded from below by the corresponding reduction of the von Neumann entropy
	\begin{equation}
	\label{eq:entropic_bound}
	{\cal B}_{E}\equiv\frac{1}{\beta}\Delta S,
	\end{equation}
We thus refer to \eqref{eq:entropic_bound} as the {\it entropic bound}.

\subsection{Thermodynamic bound}
Recently, growing interest in the thermodynamics of quantum systems has induced a closer examination of the relation between the dissipated energy and its fluctuation in the quantum information erasure process \cite{goold15}.
By considering the probability distribution function (pdf) $P(\Delta Q)$ for the net energy dissipation during the erasure process, the positiveness of the pdf and the convexity of the Boltzmann factor for the dissipated energy~\cite{convexity} allow using the well-known Jensen's inequality to have the relation
	\begin{equation}\label{eq:thermodynamic_bound}
	\beta\langle\Delta Q\rangle\geq-\ln\langle e^{-\beta\Delta Q}\rangle,
	\end{equation}
where the statistical average is taken over the pdf as $\langle e^{-\beta\Delta Q}\rangle=\int^{\infty}_{-\infty}d\Delta Q e^{-\beta\Delta Q}P(\Delta Q)$.
The inequality implies that the dissipated energy is bounded from below by the quantity 
	\begin{equation}\label{eq:thermodynamic_bound}
	{\cal B}_{T}\equiv-\frac{1}{\beta}\ln\langle e^{-\beta\Delta Q}\rangle.
	\end{equation}
We thus refer to \eqref{eq:thermodynamic_bound} as the {\it thermodynamic bound}.

\section{Full-counting statistics formalism}
The dissipated energy $\langle\Delta Q\rangle$ and the thermodynamic bound ${\cal B}_{T}$ can be evaluated by using the full counting statistics (FCS) based on a two-point projective measurement of the reservoir energy $H_{R}$ \cite{guarnieri17, hashimoto19, esposito09}.
The measurement scheme is summarized as follows:
first, at $\tau=0$, we measure $H_{R}$ to obtain an outcome $E_{0}$,
secondly, during $0\leq \tau\leq t$, the system undergoes a time evolution brought by the system--reservoir coupling,
finally, at $\tau=t$, we measure $H_{R}$ once again to obtain another outcome $E_{t}$.
The net amount of dissipated energy during the time interval $t$ is therefore given by $\Delta Q=E_{t}-E_{0}$, where its sign is chosen to be positive when the energy is transferred from the system to the environment.
The statistics of $\Delta Q$ is summarized in its probability distribution function
	\begin{equation}\label{eq:pdf}
	P(\Delta Q,t)\equiv\sum_{E_{t},E_{0}}\delta[\Delta Q-(E_{t}-E_{0})]P[E_{t},E_{0}],
	\end{equation}
with the joint probabilities obtainning the measurement outcomes
	\begin{equation}\label{eq:jpd}
	P[E_{t},E_{0}]\equiv{\rm Tr}[P_{E_{t}}U(t,0)P_{E_{0}}W(0)P_{E_{0}}U^{\dagger}(t,0)P_{E_{t}}],
	\end{equation}
where $P_{E_{\tau}}$ represents the eigenprojector of $H_{R}$ associated with the eigenvalue $E_{\tau}$, $U(t,0)$ represents the unitary time evolution of the total system, and $W(0)$ is the initial state of the total system.
Cumulants of $\Delta Q$ are provided by the cumulant generating function (cgf)
	\begin{equation}\label{eq:cgf}
	\Theta(\chi,t)\equiv\ln\int^{\infty}_{-\infty}d\Delta QP(\Delta Q,t)e^{-\chi\Delta Q},
	\end{equation}
where $\chi$ is the {\it counting field} associated with $\Delta Q$, e.g., the mean value is given by the first derivative of cgf as
	\begin{equation}\label{eq:heat}
	\langle\Delta Q\rangle=\frac{\partial \Theta(\chi,t)}{\partial(-\chi)}\biggr|_{\chi=0}.
	\end{equation}
Despite the usual definition of the cgf employing the mean value of $e^{i\chi\Delta Q}$ \cite{esposito09}, here we employ $e^{-\chi\Delta Q}$.
This change enables us to make a direct connection between the cgf and the mean value of the Boltzmann factor in Eq.~\eqref{eq:thermodynamic_bound} as
	\begin{equation}
	\Theta(\beta,t)=\ln\int^{\infty}_{-\infty}d\Delta QP(\Delta Q,t)e^{-\beta\Delta Q}=\ln\langle e^{-\beta\Delta Q}\rangle.
	\end{equation}
Thus, the thermodynamic bound is directly obtained from the cgf as
	\begin{equation}
	{\cal B}_{T}(t)=-\frac{1}{\beta}\Theta(\beta,t).
	\end{equation}

The full-counting statistics provides a systematic procedure to evaluate the cgf \cite{esposito09}.
By using Eqs.~\eqref{eq:pdf} and \eqref{eq:jpd}, and introducing the evolution operator modified to include the counting field $\chi$ by $U^{(\chi)}(t,0)\equiv e^{-\chi H_{R}/2}U(t,0)e^{+\chi H_{R}/2}$ with ${\bar W}(0)\equiv\sum_{E_{0}}P_{E_{0}}W(0)P_{E_{0}}$, we have
	\begin{equation}
	\Theta(\chi,t)=\ln{\rm Tr}_{S}[\rho^{(\chi)}(t)],
	\end{equation}
where $\rho^{(\chi)}(t)\equiv{\rm Tr}_{R}[U^{(\chi)}(t,0){\bar W}(0)U^{(-\chi)-1}(t,0)]$ is the density operator including the counting field.
Note that for $\chi=0$, $\rho^{(\chi)}(t)$ reduces to the usual reduced density operator for the system S as $\rho^{(0)}(t)={\rm Tr}_{R}[W(t)]$.
Under the factorized intial condition assumed in the quantum information erasure, the time evolution of the density operator can be described by the time-convolutionless type quantum master equation \cite{kubo63,kampen74,kampen74-2,hashitsume77,shibata77,chaturvedi79,shibata80,uchiyama99,tclbook,uchiyama14}
	\begin{equation}\label{eq:qme}
	\frac{d}{dt}\rho^{(\chi)}(t)=\xi^{(\chi)}(t)\rho^{(\chi)}(t).
	\end{equation}
The superoperator $\xi^{(\chi)}(t)$ generates time evolution of $\rho^{(\chi)}(t)$.
Taking up to the second order in its cumulant expansion with respect to the system--reservoir interaction $H_{SR}$ \cite{uchiyama99}, the superoperator is given by
	\begin{equation}\label{secondorder}
	\xi^{(\chi)}(t)\rho_{S}=-\frac{i}{\hbar}[H_{S},\rho_{S}]+K_{2}^{(\chi)}(t)\rho_{S},
	\end{equation}
with
	\begin{equation}\label{dissipationter}
	K_2^{(\chi)}(t)\rho_{S}\equiv-\frac{1}{\hbar^2}\int^{t}_{0}d\tau\Tr_{R}
	[H_{SR},[H_{SR}(-\tau),\rho_{S}\otimes
	\rho^{{\rm eq}}_{R}]_{\chi}]_{\chi},
	\end{equation}
where $H_{SR}(t)\equiv e^{i(H_{S}+H_{R})t/\hbar}H_{SR}e^{-i(H_{S}+H_{R})t/\hbar}$, and $[X,Y]_{\chi}\equiv X^{(\chi)}Y-YX^{(-\chi)}$ with $X^{(\chi)}\equiv e^{-\chi H_{R}/2}Xe^{+\chi H_{R}/2}$.
We note that the familiar master equation describing the time evolution of the usual density operator is recovered by taking $\chi=0$ on Eq.~(\ref{eq:qme}).

With these formalisms, the mean value of the dissipated energy $\langle\Delta Q\rangle$, the entropic bound ${\cal B}_{E}$ and the thermodynamic bound ${\cal B}_{T}$ are respectively expressed as
	\begin{equation}\label{eq:heat}
	\langle\Delta Q\rangle=\int^{t}_{0}{\rm Tr}_{S}\Biggr[
	\frac{\partial\xi^{(\chi)}(\tau)}{\partial(-\chi)}\biggr|_{\chi=0}
	\rho^{(0)}_{S}(\tau)\Biggr]d\tau,
	\end{equation}
(see Ref.~\cite{uchiyama14} for details)
	\begin{equation}\label{eq:entropic}
	{\cal B}_{E}(t)=\frac{1}{\beta}(S(\rho_{S}^{(0)}(0))-S(\rho_{{\rm S}}^{(0)}(t))),
	\end{equation}
and
	\begin{equation}\label{eq:thermo}
	{\cal B}_{T}(t)=-\frac{1}{\beta}\Theta(\beta,t)=-\frac{1}{\beta}\ln{\rm Tr}_{S}[\rho^{(\beta)}_{S}(t)].
	\end{equation}
	
\section{Spin--boson mode}
\subsection{Model}
For simplicity, we hereafter use units with $\hbar=1$.
As a working model, we consider the spin--boson model consisting of a single spin-$1/2$ system (S) and an infinitely large bosonic reservoir (R).
The Hamiltonian for the system consists of three terms $H=H_{S}+H_{R}+H_{SR}$, with
	\begin{equation}\label{eq:model1}
	H_{S}=\frac{\omega_{0}}{2}\sigma_{z},\;H_{R}=\sum_{k}\omega_{k}b_{k}^{\dagger}b_{k}
	\end{equation}
where $\sigma_{z,x}$ denote the Pauli matrices, $\omega_0$ denotes the energy difference between the excited state and the ground state of the spin, $\omega_k$ is energy of the $k$-th bosonic mode of the reservoir and $b_{k}\;(b_{k}^{\dagger})$ annihilation (creation) operator for the boson.
The bosonic reservoir is bilinearly coupled to the spin, and the interaction direction is tilted $\theta\in[0,\pi]$ from the $x$-axis
	\begin{equation}\label{eq:model2}
	H_{SR}=(\cos\theta\sigma_{x}+\sin\theta\sigma_{z})\otimes B_{R},
	\end{equation}
with $B_{R}\equiv\sum_{k}(g_{k}b_{k}^{\dagger}+g_{k}^{*}b_{k})$, where $g_k$ is the coupling strength between the system and the $k$-th bosonic mode.
By adjusting the parameter $\theta$, we can control the direction of the system--reservoir interaction.
For $\theta=0,\;\pi$, the system--reservoir interaction is transversal, thus the dynamics of the population and the coherence is decoupled as in the case of the previous study~\cite{hashimoto20}.
For $\theta=\pi/2$, the system Hamiltonian $H_{S}$ commutes with the interaction Hamiltonian $H_{SR}$, thus the system energy is invariant.
In the sense that the dynamics include only phase relaxation, this case corresponds to pure-dephasing.

We note that the above-presented model is equivalent to a system consisting of a single spin subjected to a tilted magnetic field and the bosonic reservoir as shown in Appendix~\ref{app:tilted-magnetic-field}. 
Even adjusting the system--reservoir interaction being challenging to realize experimentally, applying the tilted magnetic field to the spin may be much easier.

\subsection{The Bloch vector representation}
By assuming a sufficiently weak system--reservoir coupling, we employ the 2nd-order TCL master equation Eqs.~\eqref{eq:qme}--\eqref{dissipationter} to describe the thermalization process of the system.
In this paper, we focus on the interplay between the dynamics of the population and the coherence.
For this purpose, it is convenient to introduce the Bloch vector representation of the density operator because its $x,y$- and $z$-components respectively represent coherence and population.

In the presence of the counting field, the density operator of the spin $\rho^{(\chi)}_{S}(t)$ is represented by the Bloch vector including the counting field ${\bm v}^{(\chi)}(t)=(v_{x}^{(\chi)}(t),v_{y}^{(\chi)}(t),v_{z}^{(\chi)}(t),v_{0}^{(\chi)}(t))^{{\rm T}}$ with $v_{\mu}^{(\chi)}(t)\equiv{\rm Tr}_{S}[\sigma_{\mu}\rho^{(\chi)}_{S}(t)]$ ($\mu=x, y, z, 0$), where $\sigma_{0}\equiv I$ is the identity operator.
The fourth component is required because the unity of the trace of $\rho^{(\chi)}_{S}(t)$ is not held for $\chi\not=0$.
Because the density operator $\rho^{(\chi)}_{S}(t)$ is reduced to the ordinary density operator for $\chi=0$, the Bloch vector is also reduced to the ordinary Bloch vector as ${\bm v}^{(0)}(t)=(v_{x}^{(0)}(t),v_{y}^{(0)}(t),v_{z}^{(0)}(t),1)^{{\rm T}}$.

Using the Bloch vector representation, the master equation \eqref{eq:qme} is cast into the form as
	\begin{equation}\label{eq:blocheq}
	\frac{d}{dt}{\bm v}^{(\chi)}(t)=G(t){\bm v}^{(\chi)}(t),
	\end{equation}
with the $4\times4$ matrix
	\begin{equation}\label{eq:bloch_matrix}
	G^{(\chi)}(t)=\left(\begin{array}{c|c}
	A^{(\chi)}_{11}(t) & A^{(\chi)}_{12}(t)\\ \hline
	A^{(\chi)}_{21}(t) & A^{(\chi)}_{22}(t)
	\end{array}\right),
	\end{equation}
where $A^{(\chi)}_{ij}(t)$ ($i,j=1,2$) are $2\times2$ block matrices, whose lengthy expressions are summarized in Appendix~\ref{app:bloch}.
Among the four blocks, the diagonal blocks $A^{(\chi)}_{11}(t)$ and $A^{(\chi)}_{22}(t)$ describe time-evolution of coherence and population, respectively.
The off-diagonal blocks $A^{(\chi)}_{12}(t)$ and $A^{(\chi)}_{21}(t)$ describe coupling between the coherence and the population.
Importantly, the off-diagonal blocks $A^{(\chi)}_{12}$ and $A^{(\chi)}_{21}$, Eqs.~\eqref{eq:app:offdiagonal1}, \eqref{eq:app:offdiagonal2}, are proportional to $\sin2\theta$, thus they vanish for $\theta=0,\pi$ as well as for $\theta=\pi/2$.
In this case, the time evolutions of the population and the coherence are decoupled.
Otherwise, for $\theta\not=0,\pi/2,\pi$, the quantum coherence influences the population dynamics.
We also note that, for $\theta=\pi/2$, the diagonal block $A_{22}^{(\chi)}(t)$ vanishes for $\chi=0,\;\beta$ indicating invariance of $v_{z}^{(0)}(t)$, $v_{z}^{(\beta)}(t)$ and $v_{0}^{(\beta)}(t)$, (see Eq.~\eqref{eq:purediffu1} and \eqref{eq:purediffu2} in Appendix~\ref{app:bloch}).
Physically, the dynamics involve only dephasing but no population (energy) relaxation.

In terms of the Bloch vector, the bounds are formally expressed as
	\begin{eqnarray}\label{eq:enbound_bloch}
	{\cal B}_{E}(t)=&&-\ln\sqrt{1-|{\bm v}(0)|^2}
	-|{\bm v}(0)|{\rm artanh}|{\bm v}(0)|\nonumber\\
	&&+\ln\sqrt{1-|{\bm v}(t)|^2}+|{\bm v}(t)|{\rm artanh}|{\bm v}(t)|,\nonumber\\
	\end{eqnarray}
with $|{\bm v}(t)|\equiv\sqrt{(v^{(0)}_{x}(t))^2+(v^{(0)}_{y}(t))^2 +(v^{(0)}_{z}(t))^2}$, and
	\begin{equation}\label{eq:thbound_bloch}
	{\cal B}_{T}(t)=-\ln(v_{0}^{(\beta)}(t)).
	\end{equation}
Since the cumulant generating function is expressed as $\Theta(\eta,t)=\ln v_{0}^{(\eta)}(t)$, the mean dissipated energy, Eq.~(\ref{eq:heat}), is rewritten as
	\begin{equation}\label{eq:heat_bloch}
	\langle\Delta Q\rangle
	=\frac{\partial v_{0}^{(\chi)}(t)}{\partial(-\chi)} \biggr|_{\chi=0}.
	\end{equation}
From these formal expressions, we find that both of the thermodynamic bound ${\cal B}_{T}(t)$ and the mean dissipated energy $\langle\Delta Q\rangle$ are associated with $v_{0}^{(\chi)}(t)$.
In contrast, the entropic bound depends on the components $v_{x,y,z}^{(\chi)}(t)$.

\section{Relative tightness of the bounds}
We examine the relative tightness of the bound ${\cal B}_{T,E}$ against the dissipated energy $\langle\Delta Q\rangle$ in the presence of quantum coherence.
Here, we regard a bound as tighter if the bound take a closer value to the dissipated energy.
For this purpose, we numerically evaluate the bounds and the dissipated energy using the expressions Eqs.~\eqref{eq:enbound_bloch}--\eqref{eq:heat_bloch}.
In the following numerical calculations, the time interval $t$ was taken sufficiently long as the system reached the steady-state.
To describe the system--reservoir coupling, we use the Ohmic spectral density with the exponential cutoff $J(\omega)\equiv\sum_{k}|g_{k}|^2\delta(\omega-\omega_{k})=\lambda\omega\exp[-\omega/\Omega]$, where $\lambda$ is the coupling strength and $\Omega$ is the cutoff frequency.
We choose $\omega_0$ as the frequency unit for the numerical calculations.

\subsection{Dependence on initial state}\label{sec:init}
	\begin{figure}[h]
	\centering
	\includegraphics[keepaspectratio, scale=0.7,angle=0]{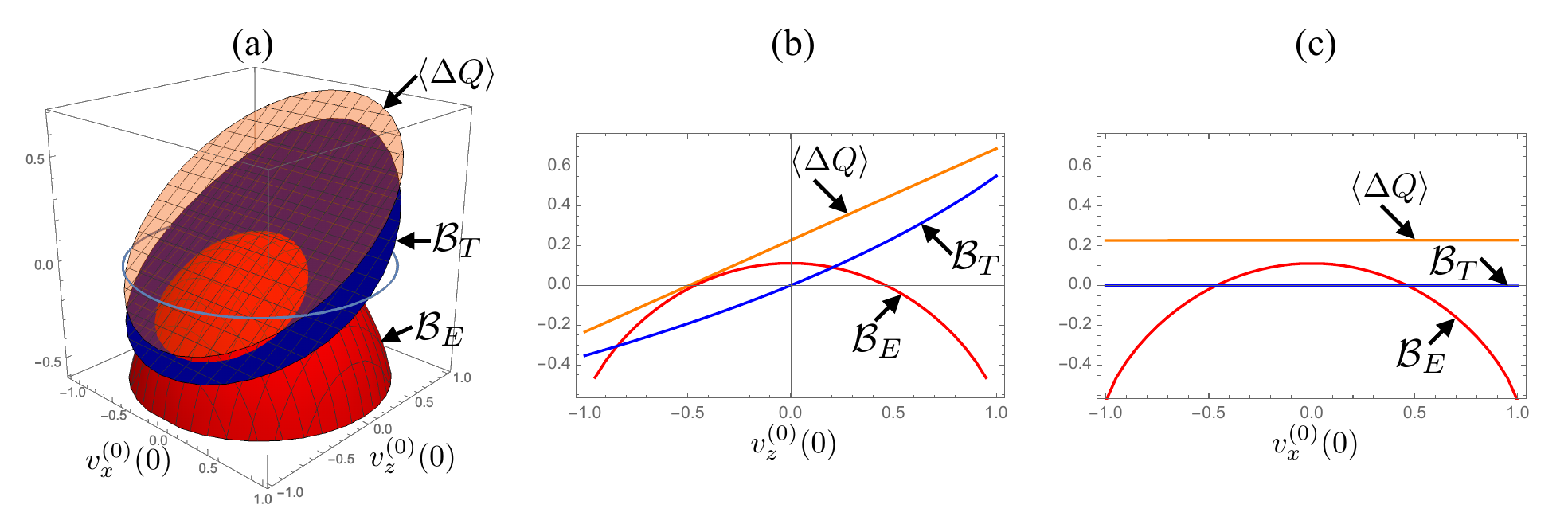}
	\vspace{-10pt}
	\caption{
	Dependences of the energy $\langle\Delta Q\rangle$ and the bounds ${\cal B}_{T,E}$ on the initial state of the system for $\theta=\pi/4$. The initial condition is chosen by changing $v_{x}^{(0)}(0)$ and $v_{z}^{(0)}(0)$ while fixing $v_{y}(0)=0$. (a) 3D plot of $\langle\Delta Q\rangle$ (orange surface), ${\cal B}_{T}$ (blue surface) and ${\cal B}_{E}$ (red surface) with respect to $v_{x}(0)$ and $v_{z}(0)$. The purple circle indicates the surface of Bloch sphere with $v_{y}^{(0)}(0)=0$. (b) cross-section of the 3D plot at $v_{x}^{(0)}(0)=0$ plotted with respect to $v_{z}^{(0)}(0)$. (c) cross-section at $v_{z}^{(0)}(0)=0$ plotted with respect to $v_{x}^{(0)}(0)$. For the numerical calculations, we set the parameters to $\lambda=0.01$, $\Omega=1$, and $\beta=1$.
	}
	\label{fig:initial1}
	\end{figure}
Let us first examine the initial state dependence of the relative tightness in the case where the time evolutions of the population and the coherence are coupled.
In Fig.~\ref{fig:initial1}, we set $\theta=\pi/4$ and plot values of the bounds and the dissipated energy for systematically changed initial states.
In panel (a), we show a 3D plot of the dissipated energy $\langle\Delta Q\rangle$ (orange surface), the thermodynamic bound ${\cal B}_{T}$ (blue surface) and the entropic bound ${\cal B}_{E}$ (red surface) with respect to $v_{z}^{(0)}(0)$ and $v_{x}^{(0)}(0)$ while setting $v_{y}^{(0)}(0)=0$.
The panels (b) and (c) show cross-sections of the panel (a) at $v_{x}^{(0)}(0)=0$ and at $v_{z}^{(0)}(0)=0$, respectively.
The figures show that both bounds are always located below the dissipated energy, meaning that both quantities properly bind from below the dissipated energy.

In the figures, we see the following difference: the dissipated energy $\langle\Delta Q\rangle$ and the thermodynamic bound ${\cal B}_{T}$ monotonically decrease as $v_{z}(0)$ decreases but they are independent of $v_{x}^{(0)}(0)$, while the entropic bound ${\cal B}_{E}$ depends isotropically on both $v_{x}^{(0)}(0)$ and $v_{z}^{(0)}(0)$ and decreases for growing $|{\bm v}(0)|$.
Because of the difference, the relative tightness of the bounds exhibits a clear boundary where the tightness switches; see the region where the red surface intersects with the blue surface.
As a consequence, the entropic bound serves as the tighter bound if the initial state is sufficiently mixed as it is located near the center of the Bloch sphere; in contrast, the thermodynamic bound is tighter if the purity of the initial state is sufficiently high as it is located near the surface of the Bloch sphere.
These qualitative features of the bounds are in agreement with the case for $\theta=0$, where the time evolutions of the population and the coherence are decoupled, studied in the previous study in Ref.~\cite{hashimoto20}, indicating that the above-summarized dependencies of the bounds on the initial state generically hold regardless of whether the dynamics influenced from the quantum coherence or not.

	\begin{figure}[h]
	\centering
	\includegraphics[keepaspectratio, scale=0.7,angle=0]{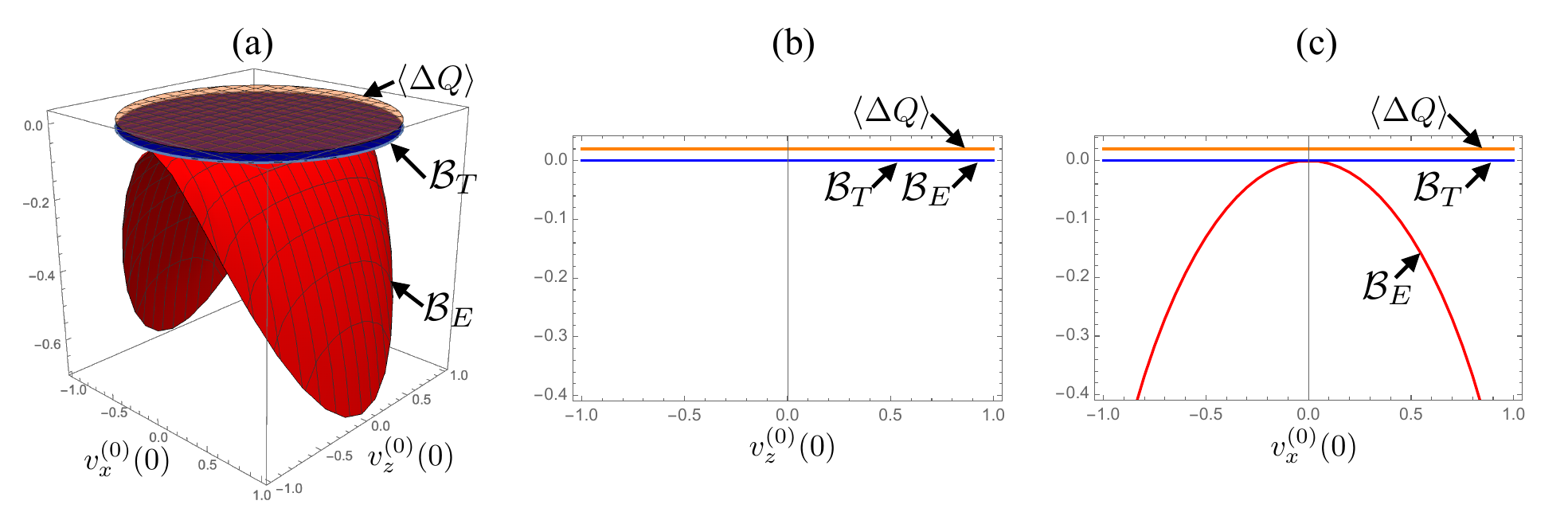}
	\vspace{-10pt}
	\caption{
	Dependences of the energy $\langle\Delta Q\rangle$ and the bounds ${\cal B}_{T,E}$ on the initial state of the system for $\theta=\pi/2$. (a) 3D plot of $\langle\Delta Q\rangle$ (orange surface), ${\cal B}_{T}$ (blue surface) and ${\cal B}_{E}$ (red surface) with respect to $v_{x}(0)$ and $v_{z}(0)$. (b) cross-section of the 3D plot at $v_{x}^{(0)}(0)=0$ plotted with respect to $v_{z}^{(0)}(0)=0$. (c) cross-section at $v_{z}^{(0)}(0)=0$ plotted with respect to $v_{x}^{(0)}(0)=0$. For the numerical calculations, we set the parameters to $\lambda=0.01$, $\Omega=1$, and $\beta=1$ (same as in Fig.~\ref{fig:initial1}).
	}
	\label{fig:initial2}
	\end{figure}
The only exception is the pure-dephasing case, $\theta=\pi/2$, presented in Fig.~\ref{fig:initial2}.
In this case, both of the dissipated energy $\langle\Delta Q\rangle$ and the thermodynamic bound ${\cal B}_{T}$ are constant for arbitrary $v^{(0)}_{x}(0)$ and $v^{(0)}_{z}(0)$, while the entropic bound ${\cal B}_{E}$ coincides with ${\cal B}_{T}$ on $v^{(0)}_{z}(0)$ axis and it decreases as $|v_{x}^{(0)}(0)|$ increases.
We also see that the dissipated energy takes a non-zero ($\approx0.02$) positive value, indicating that a certain amount of energy dissipation to the reservoir occurs regardless of the initial state.

The constant energy dissipation can be understood from the uncorrelated initial state $\rho_{{\rm tot}}(0)=\rho_{S}(0)\otimes\rho^{{\rm eq}}_{R}$ (see the third prerequisite of the quantum information erasure protocol in Sec.~\ref{sec:protocol}) and the invariance of the system energy.
Since the total system is prepared in the uncorrelated state, the exchange of energy driven by the interaction ${\cal H}_{SR}$ creates a system--reservoir correlation, which results in an attractive force.
The creation of the attractive force corresponds to the withdrawal of certain energy from the system--reservoir interaction, and the energy dissipates to the reservoir because the system energy is invariant in the pure-dephasing case. 

The thermodynamics bound is constantly zero.
It is a direct consequence of the invariance of the trace $v_{0}^{(\beta)}(t)$;
because the density operator initially coincides with the ordinary density operator $\rho_{S}^{(\beta)}(0)=\rho_{S}^{(0)}(0)$, unity of the trace hold for arbitrary $t>0$.
The behavior of the entropic bound can be understood from the pure-dephasing character of the system dynamics;
since the dynamic involves only dephasing, the states located on the $v_{z}^{(0)}(0)$ axis are invariant over time, and the states with $v_{x}^{(0)}(0)\not=0$ suffer dephasing.
Regarding the relative tightness, both bounds coincide for initial states with $v_{x}^{(0)}(0)=0$, while the thermodynamic bound serves as a tighter bound for arbitrary initial states with $v_{x}^{(0)}(0)\not=0$.

\subsection{Dependence on quantum coherence}
	\begin{figure}[h]
	\centering
	\includegraphics[keepaspectratio, scale=0.7,angle=0]{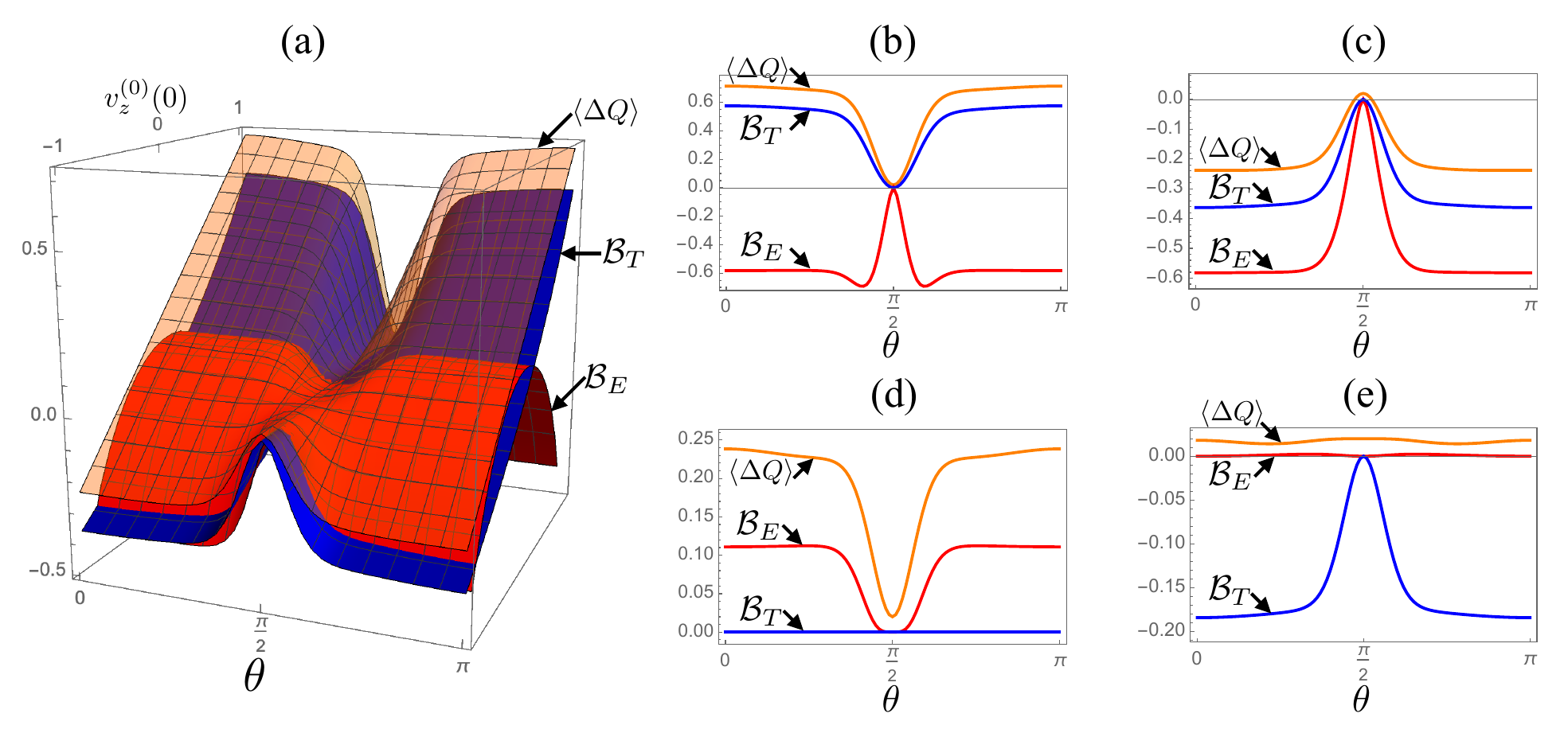}
	\vspace{-10pt}
	\caption{
	Dependences of the energy $\langle\Delta Q\rangle$ and the bounds ${\cal B}_{T,E}$ on the coherence parameter $\theta$ and the initial population $v_{z}^{(0)}(0)$ with setting $v_{x}^{(0)}(0)=v_{y}^{(0)}(0)=0$. The panel (a) show 3D plot of $\langle\Delta Q\rangle$ (orange surface), ${\cal B}_{T}$ (blue surface) and ${\cal B}_{E}$ (red surface). Panels (b) and (c): cross-sections of the 3D plot for two pure initial states (b) $v_{z}^{(0)}(0)=1$ and (c) $v_{z}^{(0)}(0)=-1$. Panels (d) and (e): cross-sections of the 3D plot corresponding to thermal initial states $\rho_{S}^{(0)}(0)=\exp[-\beta_{S}H_{S}]/{\rm Tr}_{S}[\exp[-\beta_{S}H_{S}]]$ with (d) $\beta_{S}=0$ (high temperature limit) and (e) $\beta_{S}=1(=\beta)$. For the numerical calculations, we set the parameters to $\lambda=0.01$, $\Omega=1$, and $\beta=1$ (same as in Fig.~\ref{fig:initial1}).
	}
	\label{fig:coherence}
	\end{figure}
Let us now examine the dependences of the bounds on the strength of the coherence--population coupling, controlled by the parameter $\theta$.
In Fig.~\ref{fig:coherence}, we plot values of the bounds and the dissipated energy for systematically changed coherence parameter $\theta$ and the initial population $v_{z}^{(0)}(0)$ with setting $v_{x}^{(0)}(0)=v_{y}^{(0)}(0)=0$.
In panel (a), we provide a 3D plot of the dissipated energy $\langle\Delta Q\rangle$ (orange surface), the thermodynamic bound ${\cal B}_{T}$ (blue surface) and the entropic bound ${\cal B}_{E}$ (red surface) with respect to $\theta$ and $v_{z}^{(0)}(0)$.
The panels (b)--(e) show cross-sections of the 3D plot for two pure states with (b) $v_{z}^{(0)}=1$ and (c) $v_{z}^{(0)}=-1$ as well as for two thermal mixed states, whose population is represented by $v_{z}^{(0)}(0)=(\exp[-\beta_{S}\omega_{0}/2]-\exp[+\beta_{S}\omega_{0}/2])/(\exp[-\beta_{S}\omega_{0}/2]+\exp[+\beta_{S}\omega_{0}/2])$ with an effective inverse temperature $\beta_{S}$,  with (d) $\beta_{S}=0$ and (e) $\beta_{S}=1(=\beta)$.

From the figures, we find that the dissipated energy and the bounds are insensitive to $\theta$ except for $\theta\approx\pi/2$.
Regarding the relative tightness of the bounds, the entropic bound serves as a tighter bound for most thermal initial states with positive effective temperatures, corresponding to the region with $v_{z}^{(0)}(0)<0$, while the thermodynamic bound is tighter for pure initial states, $v_{z}^{(0)}(0)=+1$ (panel (b)) and $v_{z}^{(0)}(0)=-1$ (panel (c)), or most states with negative temperatures $\beta_{S}<0$, corresponding to the region with $v_{z}^{(0)}(0)>0$.
The drastic changes in the quantities in the vicinity of $\theta=\pi/2$ are caused by the pure-dephasing character of the system dynamics.
In the region, the system dynamics are dominated by dephasing and the population relaxation only gives a minor contribution, thus the quantities rapidly change to recover their behavior at $\theta=\pi/2$ presented in Fig.~\ref{fig:initial2}.

\section{Conclusions and discussions}
In the present paper, we have examined the properties of two lower bounds for energy dissipation associated with Reeb and Wolf's quantum information erasure under the influence of quantum coherence.
As a working model, we considered a single spin-$1/2$ and a bosonic reservoir with a tilted system--reservoir interaction direction, where we could control the coupling between the dynamics of the population and the coherence by adjusting the angle of the interaction direction $\theta$.

By setting the angle to be switching on the population--coherence coupling, we found that the bounds show the following trends:
the entropic bound serves as the tighter bound if the initial state is sufficiently mixed;
while, if the purity of the initial state is sufficiently high, the thermodynamic bound is tighter.
These trends are in agreement with the case in which the population and the coherence are decoupled~\cite{hashimoto20}.
It indicates that these dependencies of the bounds on the initial state generically hold regardless of whether the influence of the quantum coherence is present or not.
Indeed, we showed that the bounds and the dissipated energy are insensitive to changing the angle for most values of $\theta$.

The only exception is the case where the angle of the interaction direction is set to $\theta=\pi/2$.
In this case, the dynamics involve only dephasing, but no energy relaxation occurs.
As a result, both dissipated energy and the thermodynamic bound are independent of the initial state, while the entropic bound decreases as the initial coherence increases.
Regarding the relative tightness, the two bounds coincide when the initial coherence is zero; otherwise, the thermodynamic bound serves as the tighter bound.

Apart from the quantum coherence between the ground state and the excited state of the spin, the constant energy dissipation caused by the system--reservoir interaction (see Fig.~\ref{fig:initial2} and its explanations in the main text) indicates that the coherence between the system and the reservoir is also a non-negligible source of energetic cost in quantum information erasure.
Even it clearly appears in the pure-dephasing case, the energy dissipation due to the system--reservoir interaction always occurs within Reeb and Wolf's framework of the erasure.
This is because the creation of the system--reservoir correlation in the course of the erasure process is inevitable for the factorized initial state assumed in its third prerequisite.
Indeed, in Ref.~\cite{shirai21}, the authors showed that the system--reservoir interaction gives a non-negligible influence on the performance of a quantum Otto engine, especially in the non-Markovian scenario.
The inclusion of the energetic cost for erasure caused by the interaction needs further investigation.

In this paper, we have studied the quantum information erasure stored in the single spin system by contacting a bosonic reservoir and bringing the spin to its steady state.
Even such a setup is universally found in energy dissipation in open systems, it is rather minor as the information erasure protocol in quantum information processing.
Indeed, recent studies~\cite{miller20, vu22} focus on the erasure by externally controlling the spin and bringing it to the ground state.
Particularly, in Ref.~\cite{vu22}, it is shown that the external driving creates quantum coherence and it inevitably causes additional energetic cost, thus it may affect the relative performance of the bounds.
Extension of this work to include the effect of the external driving is also left for future investigations.

While we have considered in this paper the spin-$1/2$ interacting with the infinite bosonic reservoir describing the surrounding radiation field or phonon field, another important source of dissipation is the coupling with surrounding spins~\cite{breuer04,cucchietti05,camalet07,segal14,mirza21}.
Indeed, in actual implementations of the qubit, such as the semiconductor quantum dot~\cite{taylor03,wu10,jing18} or the nitrogen-vacancy center in diamond~\cite{ivady20,kwiatkowski20}, coupling with surrounding nuclear spins causes energy dissipation and decoherence.
In some studies~\cite{wu10,jing18}, it is pointed out that an electron spin interacting with the collective spin reservoir shows a strong non-Markovian feature and long-lived quantum coherence.
Since these features of the spin reservoir affect the quantum information erasure, there are several efforts to
study the erasure via a finite spin reservoir~\cite{vaccaro11,croucher21}.
Thus, it is worthwhile to extend the present study to the spin reservoir case.

\appendix
\section{The Bloch equation including the counting field}\label{app:bloch}
The block matrixes $A^{(\chi)}_{ij}(t)$ in the Bloch vector representation of the master equation, Eq.(\ref{eq:blocheq}), are expressed as
	\begin{equation}
	A^{(\chi)}_{11}(t)=\begin{pmatrix}
	a_{-}^{+(\chi)}(t)\cos^{2}\theta+c_{+}^{+(\chi)}\sin^{2}\theta & -\omega_{0}+b_{-}^{+(\chi)}(t)\cos^{2}\theta\\
	\omega_{0}-b_{+}^{+(\chi)}(t)\cos^{2}\theta & a_{+}^{+(\chi)}(t)\cos^{2}\theta+c_{+}^{+(\chi)}(t)\sin^{2}\theta
	\end{pmatrix},
	\end{equation}
	\begin{equation}\label{eq:app:offdiagonal1}
	A^{(\chi)}_{12}(t)=\frac{1}{2}\sin2\theta\begin{pmatrix}
	c_{-}^{+(\chi)}(t)-a_{+}^{-(\chi)}(t)) & -ib_{+}^{-(\chi)}(t)\\
	-b_{+}^{+(\chi)}(t) & i(a_{+}^{-(\chi)}(t)-c_{+}^{-(\chi)}(t))
	\end{pmatrix},
	\end{equation}
	\begin{equation}\label{eq:app:offdiagonal2}
	A^{(\chi)}_{21}(t)=\frac{1}{2}\sin2\theta\begin{pmatrix}
	a_{-}^{+(\chi)}(t)-c_{+}^{+(\chi)}(t) & b_{-}^{+(\chi)}(t)\\
	-ib_{-}^{-(\chi)}(t) & i(a_{-}^{-(\chi)}(t)-c_{-}^{-(\chi)}(t))
	\end{pmatrix},
	\end{equation}
and
	\begin{equation}
	A^{(\chi)}_{22}(t)=\begin{pmatrix}
	a_{+}^{+(\chi)}(t)\cos^{2}\theta+c_{-}^{+(\chi)}(t)\sin^{2}\theta & ib_{+}^{-(\chi)}(t)\cos^{2}\theta\\
	ib_{-}^{-(\chi)}(t)\cos^{2}\theta & a_{-}^{+(\chi)}(t)\cos^{2}\theta+c_{-}^{+(\chi)}(t)\sin^{2}\theta
	\end{pmatrix}.
	\end{equation}
The matrix elements involve the autocorrelation function of a reservoir operator
	\begin{equation}\label{bathcorrelation}
	\langle B_{R}^{(\chi)}B_{R}^{(\chi)}(-\tau)\rangle
	\equiv{\rm Tr}_{R}[B_{R}^{(\chi)}
	B_{R}^{(\chi)}(-\tau)\rho^{{\rm eq}}_{R}],
	\end{equation}
where $B_{R}^{(\chi)}\equiv e^{-\chi H_{R}/2}B_{R}e^{+\chi H_{R}/2}$ and $B_{R}^{(\chi)}(-\tau)\equiv e^{-iH_{R}\tau}B_{R}^{(\chi)}e^{+iH_{R}\tau}$, as
	\begin{equation}
	a_{+(-)}^{\pm(\chi)}(t)\equiv-\int^{t}_{0}d\tau[h^{(\chi)}_{+(-)}(\tau)
	\pm h^{(-\chi)*}_{+(-)}(\tau)]\cos(\omega_{0}\tau),
	\end{equation}
	\begin{equation}
	b_{+(-)}^{\pm(\chi)}(t)\equiv-\int^{t}_{0}d\tau[h^{(\chi)}_{+(-)}(\tau)
	\pm h^{(-\chi)*}_{+(-)}(\tau)]\sin(\omega_{0}\tau),
	\end{equation}
	\begin{equation}
	c_{+(-)}^{\pm(\chi)}(t)\equiv-\int^{t}_{0}d\tau[h^{(\chi)}_{+(-)}(\tau)
	\pm h^{(-\chi)*}_{+(-)}(\tau)],
	\end{equation}
with
	\begin{equation}
	h^{(\eta)}_{\pm}(\tau)
	\equiv\langle B_{R}^{(\chi)}B_{R}^{(\chi)}(-\tau)\rangle
	\pm\langle B_{R}^{(-\chi)}B_{R}^{(\chi)}(-\tau)\rangle.
	\end{equation}

By setting $\theta=\pi/2$, the diagonal block $A_{22}^{\chi}(t)$ becomes
	\begin{equation}
	A^{(\chi)}_{22}(t)=\begin{pmatrix}
	c_{-}^{+(\chi)}(t) & 0\\
	0 & c_{-}^{+(\chi)}(t)
	\end{pmatrix}.
	\end{equation}
For $\chi=0$, we have $h_{-}^{(0)}(\tau)=0$ leading to $c_{-}^{+(0)}(t)=0$, thus the block vanishes as
	\begin{equation}\label{eq:purediffu1}
	A^{(0)}_{22}(t)=\begin{pmatrix}
	0 & 0\\
	0 & 0
	\end{pmatrix}.
	\end{equation}
For $\chi=\beta$, we have the equality ${\rm Tr}_{R}[b_{k}^{\dagger}b_{k}\rho^{{\rm eq}}_{R}]e^{\beta\omega_{k}}={\rm Tr}_{R}[b_{k}b_{k}^{\dagger}\rho^{{\rm eq}}_{R}]$
leading to the relations $\langle B_{R}^{(-\beta)}(-\tau)B_{R}^{(\beta)}\rangle=\langle B_{R}^{(\beta)}B_{R}^{(\beta)}(-\tau)\rangle$ and $\langle B_{R}^{(-\beta)}B_{R}^{(\beta)}(-\tau)\rangle=\langle B_{R}^{(-\beta)}(-\tau)B_{R}^{(-\beta)}\rangle$.
These relations leads to $c_{-}^{+(0)}(t)=0$, thus the block vanishes as
	\begin{equation}\label{eq:purediffu2}
	A^{(\beta)}_{22}(t)=\begin{pmatrix}
	0 & 0\\
	0 & 0
	\end{pmatrix}.
	\end{equation}

\section{A single spin subjected to a tilted magnetic field}\label{app:tilted-magnetic-field}
The working model described by Eqs.~\eqref{eq:model1}, \eqref{eq:model2} is equivalent to the system consisting of a single spin subjected to a tilted magnetic field and the bosonic reservoir.
The equivalence can be shown by applying the unitary transformation
	\begin{equation}
	U_{S}\equiv\begin{pmatrix}
	\cos\frac{\theta}{2}&\sin\frac{\theta}{2}\\
	-\sin\frac{\theta}{2}&\cos\frac{\theta}{2}
	\end{pmatrix},
	\end{equation}
to $H_{S}$ and $H_{SR}$ as
	\begin{equation}
	{\tilde H}_{S}=\frac{\omega_{0}}{2}\begin{pmatrix}
	\cos\theta & \sin\theta\\
	\sin\theta & -\cos\theta
	\end{pmatrix}\equiv
	\frac{\omega_{0}}{2}(\cos\theta{\tilde\sigma}_{z}+\sin\theta{\tilde\sigma}_{x}),
	\end{equation}
and
	\begin{equation}
	{\tilde H}_{SR}={\tilde\sigma}_{x}\otimes B_{R}.
	\end{equation}
In this view, $\omega_{0}$ represents energy splitting due to the magnetic field, and $\theta$ represents the angle of the magnetic field with respect to the $z$-axis.


\begin{thebibliography}{99}
\bibitem{convexity}
The convexity of $\Theta(\chi,t)$ with respect to $\chi$ can be directly proven from the definition (\ref{eq:cgf}).
Indeed, positivity of its second derivative with respect to $\chi$,
	\begin{equation*}
	\frac{\partial^{2}\Theta(\chi,t)}{\partial\chi^{2}}>0,
	\end{equation*}
holds for arbitrary $\chi$ because of non-negativity of the probability distribution function $P(\Delta Q,t)\geq0$.
\end{thebibliography}
\end{document}